\documentclass[pra,a4paper,twocolumn,superscriptaddress,longbibliography,nofootinbib]{revtex4-2}

\usepackage{amssymb}
\usepackage{amsmath}
\usepackage{amsfonts}
\usepackage{graphicx}
\usepackage{bm}
\usepackage{xcolor}
\usepackage{multirow}
\usepackage{comment}
\usepackage{subfigure} 
\usepackage{physics}
\usepackage{amsmath}
\usepackage{graphicx}
\usepackage{bm}
\usepackage{xcolor}
\usepackage{comment}
\usepackage{ulem}

\newcommand{\eysr}{\varepsilon_\text{YSR}}
\newcommand{\eF}{\varepsilon_\text{F}}
\newcommand{\tO}{\theta_\text{opt}} 
\newcommand{\dPair}{{\Delta_\text{pair}}}

\begin{document}

\title{Fermi-Level Pinning of Yu-Shiba-Rusinov States in a Superconductor \\ with Weakly Broken Spin-Rotational Invariance}

\author{E. S. Andriyakhina}

\affiliation{Institute of Theoretical Physics, University of Regensburg, D-93051 Regensburg, Germany}

\affiliation{Freie Universit\"at Berlin, Dahlem Center for Complex Quantum Systems and Fachbereich Physik, Arnimallee 14, Berlin, 14195, Germany}

\author{S. L. Khortsev}

\affiliation{Institute of Theoretical Physics, University of Regensburg, D-93051 Regensburg, Germany}

\author{F. Evers}

\affiliation{Institute of Theoretical Physics, University of Regensburg, D-93051 Regensburg, Germany}

\date{\today}

\begin{abstract}
As is well known, magnetic impurities adsorbed on 
superconductors, e.g. of the s-wave type, can introduce a bound gap-state (Yu-Shiba-Rusinov resonance). 
We here investigate within a minimal model how the impurity moment arranges with respect to a weak homogeneous internal magnetic field employing a fully self-consistent mean-field treatment. 
Our investigation reveals a critical window of impurity-to-substrate coupling constants, $J$. 
The width of the critical region, $\delta J$, scales like $\delta J \sim B/\Delta$, where $B$ is the magnitude of the internal field, that breaks the spin-rotation symmetry, and $\Delta$ is the bulk order parameter. While tuning $J$ through the window, the energy of the Yu-Shiba-Rusinov (YSR) resonance is pinned to the Fermi energy $\eF$ and the impurity moment rotates in a continuous fashion. 
We emphasize the pivotal role of self-consistency: In treatments ignoring self-consistency, the critical window adopts zero width, $\delta J=0$; consequently, there is no pinning of the YSR-resonance to $\eF$, the impurity orientation jumps and therefore this orientation cannot be controlled continuously by fine-tuning the coupling $J$. 
In this sense, our study highlights the significance of self-consistency for understanding intricate magnetic interactions between superconductive materials and Shiba chains.
\end{abstract}

\maketitle
\section{Introduction\label{Sec:Intro}}

The study and control of magnetic impurities in s-wave superconductors have recently gained renewed interest. Although Yu-Shiba-Rusinov (YSR) in-gap states were first predicted decades ago \cite{Yu1965,Shiba1968,Rusinov1968} and observed around 30 years ago \cite{Yazdani1997}, their unique properties and potential for technological applications continue to captivate researchers.

YSR states are now observed in a variety of superconducting systems \cite{Ruby2015,Menard2015,Yang2020,Liebhaber2020,Bode2023,Ji2008,Ruby2016,Huang2020Nat,Huang2021,Kuster2021,Trahms2023,Kezilebieke2019,Farinacci2018,Huang2020,Kezilebieke2018,Malavolti2018,Ruby2018,Friedrich2021}. These states arise from local magnetic moments introduced by adsorbed atoms or molecules and exhibit intriguing symmetries \cite{Menard2015,Yang2020,Liebhaber2020,Bode2023}, multiple bound states \cite{Ji2008,Ruby2016,Trahms2023}, and the potential for quantum phase transitions (QPT) \cite{Farinacci2018,Huang2020}. They are being actively explored for quantum computing applications, including qubit formation and the realization of Majorana fermions \cite{Choy2011,Martin2012,NadjPerge2013,Braunecker2013,Klinovaja2013,PVazifeh2013,Pientka2013,Pientka2014,Li2014,Reis2014,Rontynen2015,Menard2015,NadjPerge2014,Ruby2015PRLNov,Pawlak2016,Schneider2022}.

\begin{figure}[t]
\includegraphics[width=0.45\textwidth]{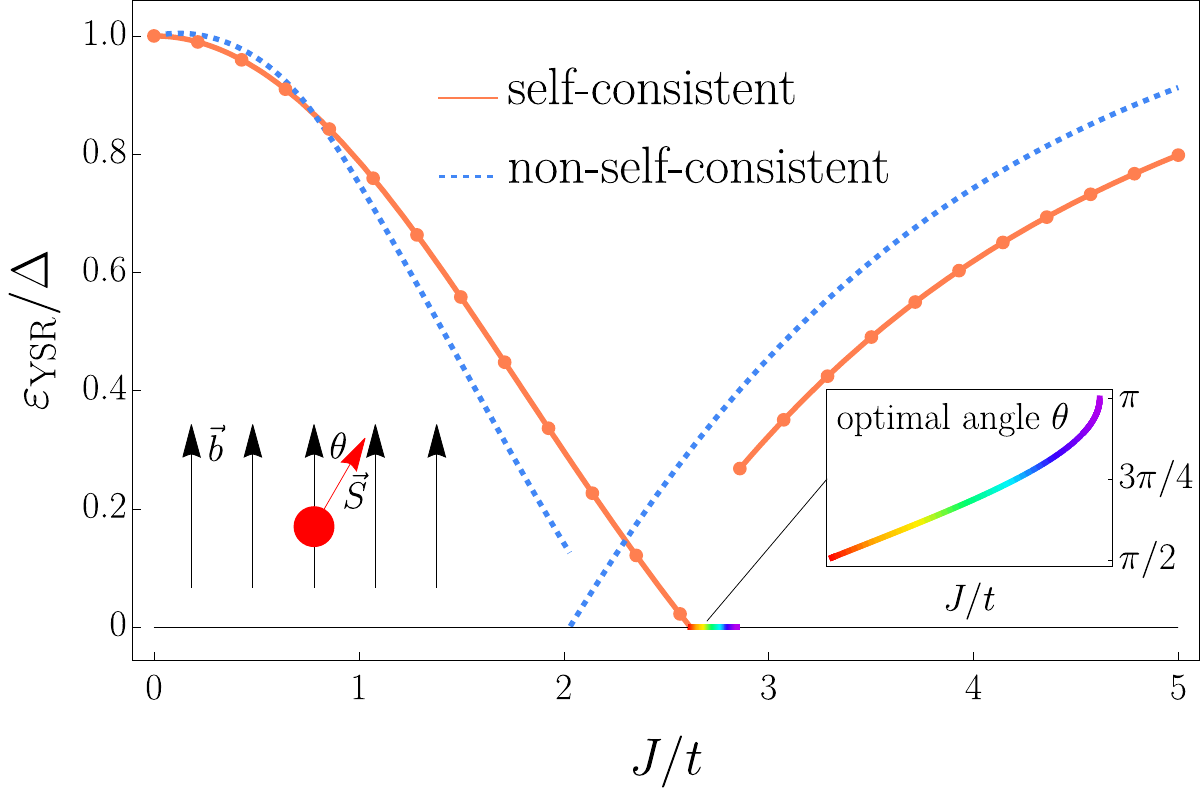}
\caption{Resonance energy of the Yu-Shiba-Rusinov state obtained in units of the bulk superconducting gap $\Delta$; self-consistent (solid trace and scattered points) and non-self-consistent calculation (dashed) are shown as functions of the exchange coupling $J$, c.f. Eq.~\eqref{eq:H_imp}. 
\textit{Insets}: The optimal angle $\theta$ between the magnetic field axis and the impurity's moment, c.f. Eq.~\eqref{eq:S_theta_phi}, as a function of $J$ through the critical region. (Parameters: $U=2.5t$, $n = 1.0$, $B = 0.05t$, and $V_{0} = 0.0t$ and $t$ the hopping amplitude.)}
    \label{fig:YSR_J}
\end{figure}

This work investigates the impact of self-consistency on the mean-field treatment of YSR states. While many studies have examined various parametric dependences \cite{Shiba1973,Salkola1997,Flatte1997,Schlottmann1976,Huang2021}, self-consistency is often neglected, with results typically valid only for weak coupling. We test the limits of this assumption by analyzing a minimal model: 
a single magnetic moment in an $s$-wave superconductor; the superconductor is assumed to be exposed to a weak, time-reversal symmetry breaking perturbation, such as a small (internal) magnetic field or a weak magnetic anisotropy in the sample. If the superconducting material has weak magnetic properties, one can attribute the origin of the $B$ field to magnetic anisotropy. 
Our results show that, without self-consistency, the YSR energy crosses the Fermi level only at a specific coupling value, $J_c$. However, self-consistent calculations reveal a critical phase where the resonance energy remains zero across a range of $J$ values, as illustrated in Fig.~\ref{fig:YSR_J}. Within this phase, the optimal angle $\theta_\text{opt}(J)$ between the magnetic moment and the superconductor's quantization direction continuously rotates from $\pi/2$ to $\pi$, as shown in the inset of Fig.~\ref{fig:YSR_J}. This phase is also characterized by a significant magnetization response near the impurity, which is discussed further in the section ``Self-consistent fields."

These findings are particularly relevant near criticality, where YSR states approach the Fermi energy, a key region for investigating Majorana Zero Modes \cite{NadjPerge2013,Pientka2013}.

\section{Model and Method\label{Sec:Model}}
 
We a consider the model Hamiltonian:
\begin{gather}
    \hat{H} = \hat{H}_0 + \hat{H}_{\rm I} + \hat{H}_{\rm M} + \hat{H}_{\rm imp}
    \label{e2} 
\end{gather}
where 
\begin{equation}
    \hat{H}_0 = - t \sum_{\langle ij \rangle, \sigma} \hat{c}_{i \sigma}^\dag \hat{c}_{j \sigma} + h.c. - \mu \sum_{i \sigma} \hat{n}_{i \sigma}
\end{equation}
denotes a nearest-neighbor tight-binding Hamiltonian with hopping parameter, $t$, and chemical potential $\mu$. The Fermionic operators $\hat{c}_{i \sigma}$ obey the usual anti-commutation relations:
$\{\hat{c}_{i \sigma}^\dag, \hat{c}_{j \sigma'} \} = \delta_{ij} \delta_{\sigma \sigma'}, \quad \{\hat{c}_{i \sigma}, \hat{c}_{j \sigma'} \} = 0$.
While the derivation of the basic equations and, hence, the results concerning the importance of self-consistency are applicable to any crystalline symmetry of the material, we present explicit results for hexagonal lattices. We are motivated by the prevalence of such symmetries in many experimental samples or systems that can be effectively reduced to hexagonal lattice configurations, such as MgB$_2$ or various transition metal dichalcogenides. However, the latter examples require more attention due to occasional charge density wave states or strong spin-orbit effects. Our letter focuses on the impact of self-consistency on a minimal model, but we acknowledge that there may be intricacies that are material-specific. Yet, we believe that the physics of the importance of self-consistency is well captured in our results and can therefore be straightforwardly extended to more intricate cases.

The Hubbard term in Eq.~\eqref{e2}, 
\begin{equation}
    \hat{H}_{\rm I} = - U \sum_i \hat{n}_{i \uparrow} \hat{n}_{i \downarrow},
    \label{e3} 
\end{equation}
embodies the onsite attractive ($U > 0$) interaction where $\hat{n}_{i \sigma}  = \hat{c}_{i \sigma}^\dag \hat{c}_{i \sigma}$ is the number operator, depicting the electron occupancy at site $i$ with spin $\sigma$.

We further incorporate the internal magnetic field introducing a Zeeman-term
\begin{equation}
    \hat{H}_{\rm M} = - B \sum_{i, \sigma\sigma'} \hat{c}_{i\sigma}^\dag (\bm{\sigma} \cdot \bm{b})_{\sigma \sigma'} \hat{c}_{i\sigma'},
\end{equation}
which breaks spin-rotation symmetry; here  
$\bm{b} = (0, 0, 1)$, and $B$ represents the magnetic field strength. We once again emphasize that in our model, the magnetic field does not couple to the magnetic impurity and further wish to clarify that $\bm{B}$ can be thought of as a source of \textit{internal} spin-rotation symmetry breaking in the substrate.
Finally, the magnetic impurity is implemented via 
\begin{equation}
    \hat{H}_{\rm imp} = - J \sum_{\sigma \sigma'} \hat{c}_{0\sigma}^\dag (\bm{\sigma} \cdot \bm{S})_{\sigma \sigma'} \hat{c}_{0 \sigma'} +  V_{0} \sum_{\sigma} \hat{c}_{0\sigma}^\dag \hat{c}_{0 \sigma} , \label{eq:H_imp}
\end{equation}
at the lattice site $\bm{r}_0$; $J$ denotes the exchange coupling constant for magnetic interactions.  A local potential with strength $V_0$ is also foreseen; it accounts, e.g., for an impurity-induced Coulomb interaction. Throughout this study, we adopt the assumption of a classical impurity spin, an approach fundamentally outlined in Ref.~\cite{Shiba1968}.

To characterize the relative orientation between the local magnetic moment $\bm{S}$ and the uniform magnetic field $\bm{b}$, we introduce a spherical parametrization for $\bm{S}$, described by the polar ($\theta$) and azimuthal ($\phi$) angles:
\begin{equation}
\bm{S} = (\cos \phi \sin \theta, \sin \phi \sin \theta, \cos \theta)^T. \label{eq:S_theta_phi}
\end{equation}
In the absence of spin-orbit coupling, the system exhibits a residual symmetry under rotations around $\bm{b}$. Since $\bm{S}$ is not equipped with a dynamics of its own, 
the azimuthal angle $\phi$ defined in Eq.~\eqref{eq:S_theta_phi} is cyclic. Therefore, subsequently we will discuss only the dependence on the polar angle $\theta$.


{\it Mean-field approximation.} In the mean-field approximation, the product of number operators on the same site simplifies to:
\begin{gather}
    \hat{n}_{i \uparrow} \hat{n}_{i \downarrow} = \sum_{\sigma\sigma'} \rho_{i\bar{\sigma}\bar{\sigma}'}  \hat{c}_{i \sigma}^\dag \hat{c}_{i \sigma'} - \rho_{i\uparrow \uparrow} \rho_{i\downarrow \downarrow} + |\rho_{i\uparrow \downarrow}|^2 \notag \\
    + \frac{\Delta_i}{U} \hat{c}_{i \uparrow}^\dag \hat{c}_{i \downarrow}^\dag + \frac{\Delta_i^*}{U}  \hat{c}_{i \downarrow} \hat{c}_{i \uparrow} - \frac{|\Delta_i|^2}{U^2},
\end{gather}
where
\begin{gather}
    \rho_{i \sigma \sigma'} = (2\delta_{\sigma \sigma'} - 1) \langle \hat{c}_{i \sigma}^\dag \hat{c}_{i \sigma'} \rangle, \quad \Delta_i = U \langle \hat{c}_{i \downarrow} \hat{c}_{i \uparrow} \rangle.
\end{gather}
Here, $\langle \hat{c}_{i \downarrow} \hat{c}_{i \uparrow} \rangle \neq 0$ is the anomalous average that appears in the superconducting state. Using these fields, the mean-field interaction Hamiltonian becomes:
\begin{gather}\label{eq:H_I_MF}
    \hat{H}_{\rm I}^{\rm MF} = - U \sum_{i, \sigma\sigma'} \rho_{i\bar{\sigma}\bar{\sigma}'}  \hat{c}_{i \sigma}^\dag \hat{c}_{i \sigma'} - \sum_{i} (\Delta_i \hat{c}_{i \uparrow}^\dag \hat{c}_{i \downarrow}^\dag + \Delta_i^* \hat{c}_{i \downarrow} \hat{c}_{i \uparrow}),
\end{gather}
with $\bar{\uparrow} = \downarrow$ denoting spin-flip processes.


{\it Solving the self-consistency problem.} We now employ the Bogoliubov--de Gennes formalism to diagonalize the Hamiltonian via the introduction of new quasiparticle operators $(\hat{\gamma}_{n}, \hat{\gamma}_{n}^\dag)$:
\begin{gather}
    \hat{c}_{i \sigma} = \sum_{n}' (u_{n \sigma}^i \hat{\gamma}_{n} - \sigma v_{n \sigma}^{i*} \hat{\gamma}_{n}^\dag), \label{eq:BdG_transform}
\end{gather}
where only positive-energy states $\varepsilon_n \geqslant 0$ are considered (denoted using the prime symbol on the summation). The Hamiltonian in our desired basis should be diagonal:
\begin{equation}
    \hat{H}^{\rm BdG} =  E_g + \sum_{n} \varepsilon_{n} \hat{\gamma}_{n}^\dag \hat{\gamma}_{n}, 
\end{equation}
which sets up the problem for funding the transformation coefficients $u_{n \sigma}^i$ and $v_{n \sigma}^i$. Here, the ground state energy
\begin{gather}
    E_g = \sum_{i \sigma}\sum_{n}' - \varepsilon_{n} |v_{n\sigma}^i|^2 + \sum_i \frac{|\Delta_i|^2}{U} \notag \\
    + U \sum_i (\rho_{i\uparrow \uparrow} \rho_{i\downarrow \downarrow} - |\rho_{i\uparrow \downarrow}|^2).\label{eq:E_g}
\end{gather}
Adopting the transformation~\eqref{eq:BdG_transform} and $T = 0$, the previously defined fields are expressed as
\begin{gather}
    \Delta_i = \frac{U}{2} \sum_{n}' (u_{n \uparrow}^i v_{n \downarrow}^{i*} - u_{n \downarrow}^i v_{n \uparrow}^{i*}), \label{eq:sc:Delta} \\
    \rho_{i\sigma\sigma'} = (2\delta_{\sigma\sigma'} - 1)\sum_{n}' v_{n \sigma}^i v_{n \sigma'}^{i*}. \label{eq:sc:n}
\end{gather}
Further details on the BdG Hamiltonian and self-consistency are provided in \cite{Supplementary}.

In every numerical simulation conducted in this study, we explored two-dimensional hexagonal lattices at zero temperature with periodic boundary conditions, having an edge length of $L = 30$ cells. Setting $U = 2.5t$ and $n = 1.0$ yielded a gap $\Delta \simeq 0.443t$ and a density of states at the Fermi level $\pi\nu_0 = 0.382/t$. Throughout this letter, we also employ $B = 0.05t$, resulting in a shift of the continuous spectrum boundary to $\Delta(B) = 0.393t$.

\section{Theoretical considerations \label{Sec:Theory}}

From an analytical perspective, solving the self-consistency problem is notoriously challenging. Consequently, researchers often ignored variations in the field distributions $\Delta_i$ and $\rho_{i\sigma\sigma'}$, assuming that effects stemming from their inhomogeneity are marginal. Following these assumptions, we formulate in this subsection some preliminary theoretical expectations. 
Intriguingly, as we show in subsequent numerical sections, self-consistency introduces notable changes to the non-selfconsistent scenario. 


{\it Resonance energy - perturbative analysis.} Assuming homogeneous distributions of $\Delta$ and $\rho_{\sigma \sigma'}$, the analytical results for the resonance energy suggest the following behavior of the in-gap energy \cite{Rusinov1968}:
\begin{equation}
    \eysr^{(0)} = \pm \Delta \frac{1-\alpha^2+\beta^2}{\sqrt{(1-\alpha^2+\beta^2)^2+4\alpha^2}}, \label{eq:no_sc:E_YSR}
\end{equation}
where $\alpha = \pi \nu_0 J$, $\beta = \pi \nu_0 V_{0}$. The resonance energy is measured from the Fermi level $\eF$, which at $T=0$ equals $\mu$, and $\nu_0$ denotes the density of states at $\eF$.  
At specific critical valuew, $\alpha_c^{(0)}(\beta) = \pm \sqrt{1+\beta^2}$, the energy drops to zero, signaling a QPT at which the spontaneous generation of a local quasiparticle excitation becomes energetically favorable.

In order to estimate the effect of the Zeeman term, embarking on \cite{Rusinov1968}, we perform a perturbative analysis referring to the Supplemental Material \cite{Supplementary} for details.  We obtain for the leading order correction
\begin{equation}
    \delta \eysr^{(1)} = \pm B \cos \theta, \text{ with} \begin{cases}
        +, & \alpha > \alpha_c^{(0)}(\beta) \\
        -, & \alpha < \alpha_c^{(0)}(\beta)
    \end{cases}, 
    \label{eq:no_sc:E_YSR}
\end{equation}
which implies a sharp jump as the coupling $J$ crosses a critical value $\alpha_c^{(0)}(\beta)/\pi\nu_0$. 
A corresponding jump is also seen in the free energy
\begin{gather}
    \delta E_g^{(1)} = - B \cos \theta \begin{cases}
        1, & \alpha > \alpha_c^{(0)}(\beta) \\
        0, & \alpha < \alpha_c^{(0)}(\beta)
    \end{cases} . \label{eq:no_sc:delta_E_g}
\end{gather}
To rationalize the result \eqref{eq:no_sc:delta_E_g}, we stipulate that the ground state undergoes a magnetic transition, evidence of which will be presented below: at 
$\alpha > \alpha_c^{(0)}(\beta)$ the net moment exhibits a spin-$\frac{1}{2}$, while at $\alpha < \alpha_c^{(0)}(\beta)$, 
it exhibts vanishing electronic spin and no net moment\cite{Sakurai1970,Salkola1997}. Therefore, there is no linear response of the ground state energy to $B$ and $\delta E_g^{(1)}=0$ in the latter regime.


\begin{figure}[t]
\includegraphics[width=0.45\textwidth]{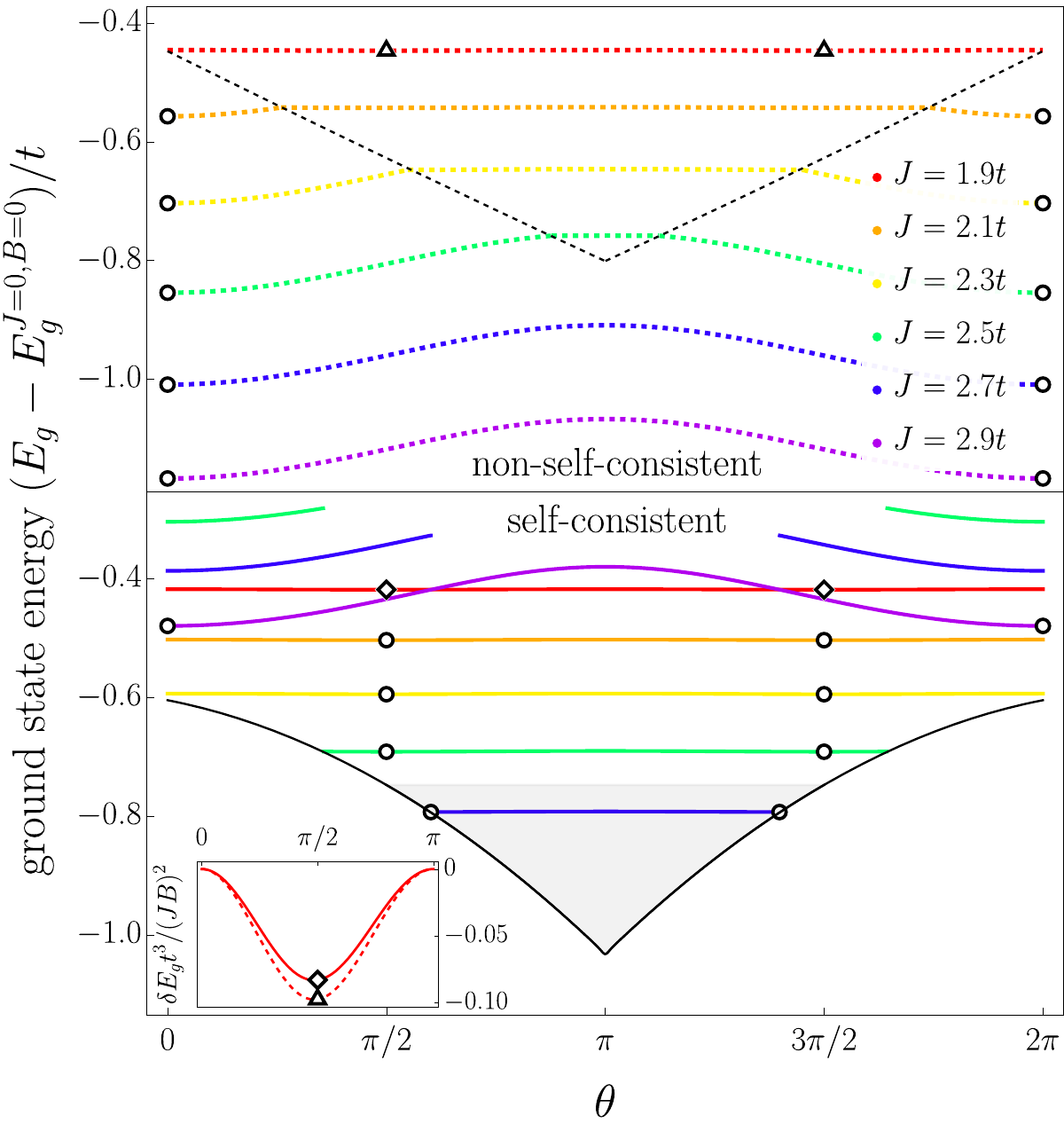}
\caption{Ground state energy, $E_g(\theta)$ over the angle $\theta$ between magnetic field and impurity moment (c.f. Fig.~\ref{fig:YSR}) for various coupling constants $J$ as obtained from numerical diagonalization. Results neglecting self-consistency (dashed traces) and from the fully self-consistent scheme (solid traces) are shown. The jump in the latter seen in the range $2.5t\lesssim J\lesssim 2.7t$ indicates a transition between spin-0 and spin-$\frac{1}{2}$ phases.
The black solid (dashed) line marks the position where $\eysr(\theta) = 0$ in (non-)self-consistent scheme, delineating the transition between phases with spin-0 and spin-$\frac{1}{2}$.  Open markers indicate the minima of the ground state energy as a function of $\theta$. Gray shading in self-consistent calculations indicates the region where the optimal angle $\theta_\text{opt}$ deviates from $0$ and $\pi/2$. 
(Parameters: $B=0.05t$) 
\textit{Inset}: Evolution of $\delta E_g = E_g(\theta) - E_g(0)$ with $\theta$ at weak coupling $J$. The plot reveals a $\cos 2\theta$ dependence, reaching its minimum at $\theta_c = \pi/2$.
(Parameters: $J=1.9t, B=0.05t$) 
}
\label{fig:E_g}
\end{figure}

{\it Resonance energy - non-self-consistent numerics.} To go beyond the perturbative limit but still discarding self-consistency, we perform the exact diagonalization numerically, keeping uniform fields $\Delta_i \equiv \Delta = 0.443t$ and $\rho_{i\sigma\sigma'} \equiv \frac{n}{2} \delta_{\sigma\sigma'} = \frac{1}{2} \delta_{\sigma\sigma'}$.
The result is shown in the upper panel of Fig.~\ref{fig:E_g}. Specifically, we display the behavior of the ground state energy as a function of the angle, $E_g(\theta)$. 

As one would have expected based on Eq.~\eqref{eq:no_sc:delta_E_g}, the $B$-dependent contribution of the ground-state energy, $\delta E_g$, at small $J\lesssim J_{c_1}$ (with $J_{c_1}\approx 1.9t$) the traces are flat (up to quadratic corrections), while at large $J\gtrsim J_{c_2}$ ($J_{c_2}\approx 2.7t$) they exhibit a cosine shape. A new feature of the numerical solution as compared to \eqref{eq:no_sc:delta_E_g} is the existence of the transition window $J_{c_1}\lesssim J \lesssim J_{c_2}$, which opens up as $B$ rises from zero. 
In this intermediate regime, for each combination of $J,V_0$ (or $\alpha,\beta$) there is an angle $\theta_c(\alpha,\beta) = \arccos\left(\varepsilon_{\rm YSR}^{(0)}/B\right)$ such that $\delta E_g(\theta)$ is a nearly-flat(in the first order in small $B$) function for $\theta_c<\theta<2\pi-\theta_c$; see the dashed triangle in 
Fig. \ref{fig:E_g}. 

The evolution of the (non-self-consistent) YSR-resonance, $\eysr$, corresponding to $\theta_c$ was already shown in Fig. \ref{fig:YSR_J} (dashed line): with increasing coupling $J$, the resonance energy splits from the continuum ($\eysr=\Delta$) and moves gradually towards the band center that, however, is ultimately reached only after a sharp jump corresponding to $\theta_c$ dropping abruptly from $\pi/2$ to zero. 

{\it Ground-state magnetization.} The minimum of $E_g(\theta)$, denoted $\theta_{\text{opt}}(\alpha,\beta)$, gives the optimal orientation of the impurity spin $\bm{S}$ relative to the internal field $\bm{B}$. For $J \gtrsim J_{c_1}$, as shown in Fig.~\ref{fig:E_g}, $\theta_{\text{opt}} = 0$, indicating that the impurity spin aligns fully with the internal field. In this regime, the local field is strong enough to break Cooper pairs, creating an electronic spin-1/2 that couples linearly to $\bm{B}$, as explained by Eq. \eqref{eq:no_sc:delta_E_g}.

For $J \lesssim J_{c_1}$, the system does not couple linearly to $B$ see Eq. \eqref{eq:no_sc:delta_E_g}, suggesting that Cooper pairs remain intact and there is no net electronic spin. As depicted in Fig.~\ref{fig:E_g}, inset, $E_g$ varies as $B^2$ for $J \lesssim J_{c_1}$, with a $\cos(2\theta)$ modulation. Here, $\theta_{\text{opt}} = \pi/2$, meaning the impurity's spin is perpendicular to the quantization axis $\bm{b}$, minimizing its impact on Cooper pair destruction.

{\it Discussion.} Any non-self-consistent implies a lack of feedback: while the eigenfunctions of ${H}_{ij}^{\rm BdG}$ gradually deform with increasing impurity strength, the coupling $\Delta$ remains unaffected. Therefore, pair breaking does not cost the energy $\sim\Delta$; consequently, the total energy of the spin-0 state shown in the upper panel of Fig. \ref{fig:E_g} is too low by an amount $\Delta$ as compared to the spin-1/2 state. This artifact is likely to invalidate a non-self-consistent description of the magnetic phase transition, so that a fully self-consistent treatment is required.

\section{Fully self-consistent numerics \label{Sec:Results}}

As we demonstrate in this section, full self-consistency changes the scenario significantly because it restores the energy cost for pair breaking.  

{\it Ground state energy.} At weak coupling ($J < J_{c_1}$), the ground state energy $E_g(\theta)$ is nearly constant, reflecting a spin-0 phase similar to results ignoring self-consistency. At strong coupling ($J > J_{c_2}$), $E_g(\theta)$ follows the familiar $\cos(\theta)$ shape, indicating a spin-1/2 phase. Unlike the non-self-consistent result, Fig. \ref{fig:E_g} (upper panel), $E_g$ evolves non-monotonously with $J$ in the lower panel: the spin-1/2 state is elevated by $\dPair \approx \Delta$, indicating the cost of pair-breaking.

This extra energy $\dPair$ causes a sharp jump in $E_g(\theta)$ within the critical window $J_{c_1} \lesssim J \lesssim J_{c_2}$ (e.g., at $J=2.5t$ and $J=2.7t$). As a result, the spin-0 state remains stable as $J$ enters the critical window, and the optimal angle $\tO(J)$ stays near $\theta = \pi/2$ instead of jumping to $\theta = 0$ as predicted without self-consistency. Interestingly, as $J$ approaches $J_{c_2}$, $\tO(J)$ first increases beyond $\pi/2$ (see inset of Fig. \ref{fig:YSR_J}), indicating that the impurity anti-aligns with the internal field to prevent Cooper-pair splitting. Only when the Zeeman energy overcomes the pair-breaking cost does the spin-1/2 ground state emerge.

\begin{figure}[t]
\includegraphics[width=0.45\textwidth]{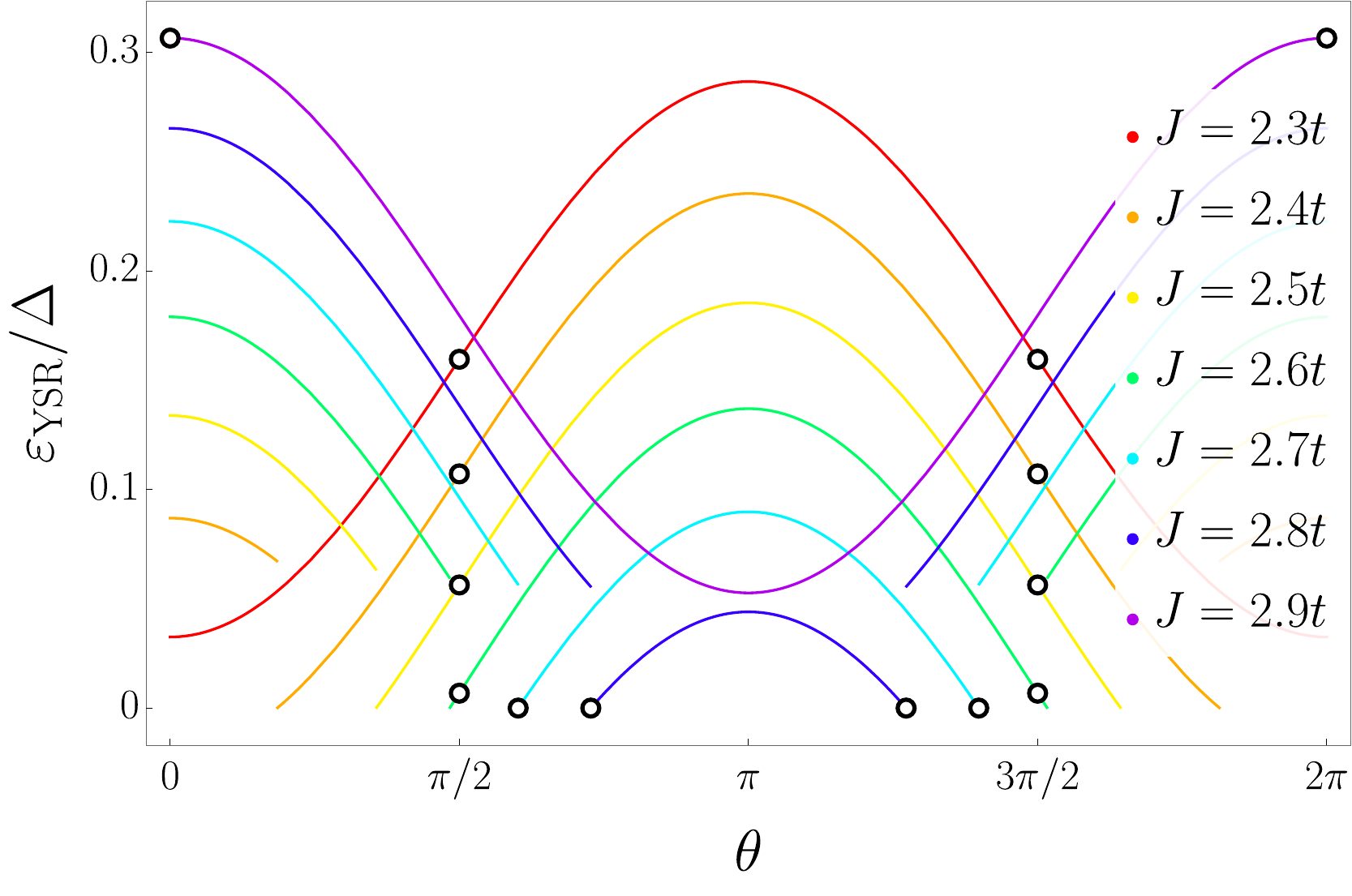}
    \caption{Evolution of the fully self-consistent in-gap energy with varying coupling $J$ and relative polar angle $\theta$. The plot illustrates the region where self-consistency drives the YSR energy to the Fermi energy (zero-energy), profoundly altering the system's state. The open circles mark the angle $\tO(J)$ of lowest total energy as extracted from Fig. \ref{fig:E_g}. (Parameters: $V_{0} = 0.0t$, $B=0.05t$)}
    \label{fig:YSR}
\end{figure}

{\it Yu-Shiba-Rusinov resonance energy.} The transition from the spin-0 to spin-1/2 state occurs through a level crossing in the mean-field framework. At weak coupling ($J < J_{c_1}$), the occupied YSR-resonance hosts a Cooper pair localized near the impurity. At very strong coupling ($J > J_{c_2}$), the Zeeman energy dominates, resulting in spin-split orbitals with one occupied (majority) and the other empty (minority).

The trace of $\eysr(\theta)$ in Fig. \ref{fig:YSR} illustrates this scenario: as $J$ reaches $J_{c_1}$, the YSR state's energy hits the Fermi energy $\eF$, signaling the transition \cite{Sakurai1970,Matsuura1977,Sakai1993}. This transition impacts local electronic states, transport, and magnetic properties \cite{Ptok2017,Huang2021,LopezBezanilla2019}. The analysis of $E_g(J)$ reveals a critical window, $J_{c_1} < J < J_{c_2}$, where $\tO(J)$ shifts from $\pi/2$ to $\pi$ before the spin-1/2 phase with $\tO=0$ emerges. As $J$ enters this window ($J > J_{c_1}$), the Cooper pair nears destruction, keeping the resonance close to $\eF$. It remains pinned to $\eF$ throughout the window until it eventually dissolves as $J$ exceeds $J_{c_2}$.

\begin{figure*}[t]
\centerline{ 
\includegraphics[width=0.8\textwidth]{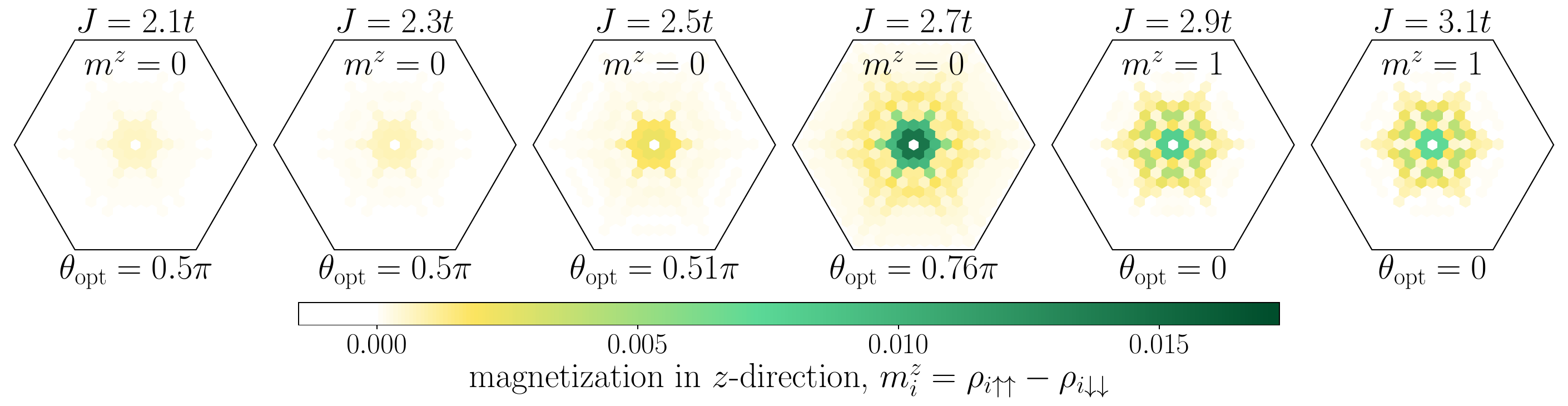}}
\caption{
Evolution of the magnetization $m^z_i = \rho_{i\uparrow\uparrow} - \rho_{i\downarrow\downarrow}$ along the $\bm{B}$-field axis with $J$ at an optimal angle $\theta_{\rm opt}(J, V_0)$. Notably, for $J < J_{c_2} \simeq 2.78t$, the total spin polarization $m^z = \sum_i m_i^z$ equals 0, while for $J > J_{c_2}$, it is 1. Additionally, at $\theta = \pi/2$, the magnetization response is minimal, as depicted in the first three figures of the row. (Parameters: $V_{0} = -0.5t$, $B = 0.05t$)}
\label{f4} 
\end{figure*}


{\it Self-consistent fields.} The spatial distribution patterns of self-consistent fields, i.e., particle densities $\rho_{\sigma\sigma'}(\bm{r})$, magnetization $m^z(\bm{r}) = \rho_{\uparrow\uparrow}(\bm{r}) - \rho_{\downarrow\downarrow}(\bm{r})$, and the order parameter $\Delta(\bm{r})$, are experimental observables. We analyze how they reflect the QPT.  For additional details, we refer to Supplemental Material \cite{Supplementary}.
In Fig. ~\ref{f4} we show the distribution of magnetization induced by the local impurity. As we observe $J$ progressing from $J < J_{c_1} \simeq 2.50t$ to $J > J_{c_2} \simeq 2.78t$, the critical region reveals itself by a distinct change in the magnetization profile and the total spin polarization $m^z = \int m^z(\bm{r})d\bm{r}$. When $J < J_{c_1}$, i.e., $\theta_{\rm opt} = \pi/2$, the magnetization response in the $\bm{b}$ direction is minimal, which plays a crucial role in the spin-0 phase. Conversely, in the region $J_{c_1} \leqslant J \leqslant J_{c_2}$, the magnetization response is anomalously large, which is evident from the figure corresponding to $J = 2.7t$ in Fig.~\ref{f4}. 


\section{Discussion\label{Sec:Discussion}}

Our analysis shows that self-consistency significantly alters the previously understood non-self-consistent picture. Near criticality, the destruction of Cooper pairs is mitigated by the reorientation of the impurities magnetic moment, where the angle $\theta_{\rm opt}$ between the internal magnetic field $\bm{B}$ and $\bm{S}$ exceeds $\pi/2$. This reorientation lowers the energy of a localized electron by reducing the effective magnetic field at the impurities site to $B + J \cos \theta_{\rm opt}$, which is less than $B$ when $\theta_{\rm opt} > \pi/2$.

While a detailed analytical solution to the self-consistent problem is beyond this letter’s scope, we can approximate the effects of self-consistency at leading order in $B/\Delta \ll 1$ with:
\begin{equation}
    \alpha_{c_1}(\beta, B) = \sqrt{1 + \beta^2}, \ \alpha_{c_2}(\beta, B) = \sqrt{1 + \beta^2} + \frac{B}{\Delta}. \label{eq:alpha_c_1-alpha_c_2}
\end{equation}
In the critical region of width approximately $B/\Delta$, we predict:
\begin{equation}
    \eysr = 0, \quad \alpha_{c_1}(\beta, B) \leqslant \alpha \leqslant \alpha_{c_2}(\beta, B).
\end{equation}
These expressions highlight that self-consistency primarily serves to mitigate Cooper pair breaking, reflected in the abrupt jump in the ground state energy $E_g$.

In real superconductors, other mechanisms like a finite density of magnetic impurities or spontaneous symmetry breaking may disrupt spin-rotation symmetry. Our results should apply in these cases as well, with the width of the critical region $B/\Delta$ being replaced by $B^*/\Delta$, where $B^*$ represents the strength of the symmetry-breaking perturbation.
In Shiba chains, self-consistency is expected to have tangible effects, potentially leading to impurity spin reorientations even without an internal field. This is crucial for impurity chains, where collective behaviors and interactions play a significant role.

Finally, we offer a brief discussion of fluctuation effects that are ignored in our discussion. (i) In the thermodynamic limit followed by $B\to 0$ time reversal symmetry is restored. One would expect a zero-temperature phase of a spin coupled to a superconducting bath with broken spin-rotational invariance to be unstable against increasing the temperature; so, the impurity does not give rise to a net magnetization. However, this does not necessarily imply irrelevance of the unstable (mean-field) fixed point for experimental observations. The situation is, in fact, quite standard in the theory of quantum phase transitions: unstable zero-temperature fixed points tend to leave their signatures at finite temperature in observables measured at finite frequencies. We believe that this observation carries over also to the present case of the YSR-impurity. 
(ii) We believe that also in quantum spin systems the mechanism described here will survive qualitatively: within a certain range of coupling constants $J$, the impurity’s spin expectation value will exhibit a bias towards being negative, aligned opposite to the internal magnetic field $B$, thereby altering the behavior of the in-gap state energy, local density of states (LDoS), and magnetic response.

We nonetheless wish to emphasize that our results have been obtained within the mean-field approximation and in the classical-spin limit. To understand the effect of quantum fluctuations, further research is necessary. For example, employing numerical renormalization group techniques, which can handle the feedback of the impurity into the bath, and treating the impurity spin quantum mechanically would be important future directions.

\section{Conclusion\label{Sec:Conclusion}}

We have examined the role of self-consistency in a minimal system with a magnetic impurity in an s-wave superconductor under a weak magnetic field, focusing on the ground state and Yu-Shiba-Rusinov states. Our analysis shows that self-consistency introduces qualitatively new effects as the dimensionless coupling, $\alpha$, approaches its critical value $\alpha_c$. Instead of a singular critical point, $\alpha_c$, a critical phase emerges where $\eysr= 0$, as seen in Fig.~\ref{fig:YSR_J}.

Future research could explore: (i) Extending our results to the quantum-impurity case and going beyond the mean-field approximation; (ii) Investigating self-consistency in YSR-spin chains and its impact on Majorana bound states and quantum phase transitions; (iii) Experimentally, detailed studies of magnetization, LDoS distributions, and YSR peak behavior with varying $J$, $V_0$, $\mu$, $B$, and $n$ would enhance our understanding. A magnetic tip in STM experiments could serve as a local magnetic impurity, with adjustable coupling strength $J$ potentially allowing phase transition observations in scanning tunneling microscopy measurements, at least, in principle.

\begin{acknowledgements}
E.S.A. and F. E. are grateful to I. Burmistrov for useful discussions and collaborations on related research. E.S.A. acknowledges useful discussions and constructive criticism from P. Brouwer, S.S. Babkin, D.E. Kiselev, and K.G. Nazaryan. The Gauss Centre for Supercomputing e.V. is acknowledged for providing computational resources on SuperMUC-NG at the Leibniz Supercomputing Centre under the Project ID pn36zo. E.S.A and F. E. acknowledge support by the Deutsche Forschungsgemeinschaft (DFG, German Research Foundation) within Project-IDs 314695032; 281653456; 430195475 – SFB 1277 (Projects No. A03 and No. IRTG). 

\end{acknowledgements}

\bibliography{bib-YSR}

\newpage
\phantom{}
\newpage
\begin{widetext}
\renewcommand{\appendixname}{}

\appendix

  \begin{center}{\large 
  Supplemental Material for \\``Fermi-Level Pinning of Yu-Shiba-Rusinov States in a Superconductor \\ with Weakly Broken Spin-Rotational Invariance" }
  \end{center}
  \maketitle

\setcounter{figure}{0}
\renewcommand{\thefigure}{S\arabic{figure}}
\setcounter{page}{1}

\setcounter{section}{0}
\renewcommand{\thesection}{\Alph{section}}
\renewcommand{\theequation}{\thesection.\arabic{equation}}

\section{Self-consistency problem \label{sec:NSC}}

This section delineates the details of the self-consistency problem we solve in the main text. After we perform the Bogoliubov--de Gennes transformation, the BdG Hamiltonian's matrix elements, in the basis $ (u_{\uparrow \sigma}, \; u_{\downarrow \sigma}, \; v_{\uparrow \sigma}, \; v_{\downarrow \sigma})^T $, become:
\begin{gather}
    {H}_{ij}^{\rm BdG} = \begin{pmatrix}
        h_{ij\uparrow\uparrow} & h_{ij\uparrow\downarrow} & 0 & \Delta_i \delta_{ij} \\
        h_{ij\downarrow\uparrow} & h_{ij\downarrow\downarrow} & -\Delta_i\delta_{ij}  & 0 \\
        0 & -\Delta_i^*\delta_{ij}  & -h_{ij\uparrow\uparrow}^* &  -h_{ij\uparrow\downarrow}^* \\
        \Delta_i^*\delta_{ij}  & 0 & -h_{ij\downarrow\uparrow}^* & -h_{ij\downarrow\downarrow}^*
    \end{pmatrix}.
\end{gather}
Here, $ h_{ij\sigma\sigma'} $ is defined as $ h_{ij\sigma\sigma'} = -t \delta_{\langle i,j \rangle} \delta_{\sigma \sigma'} - (\mu \delta_{\sigma \sigma'} + U \rho_{i\bar{\sigma} \bar{\sigma}'} + B (\bm{\sigma} \cdot \bm{b})_{\sigma\sigma'}) \delta_{ij} + (V_{0} \delta_{\sigma \sigma'} - J (\bm{\sigma} \cdot \bm{S})_{\sigma \sigma'}) \delta_{ij} \delta_{i0}$. The above equations culminate in the following eigenvalue problem for our system:
\begin{equation}
    \sum_j {H}_{ij}^{\rm BdG} \begin{pmatrix}
        u_{n\uparrow}^j \\ u_{n\downarrow}^j \\ v_{n\uparrow}^j \\ v_{n\downarrow}^j
    \end{pmatrix} = \varepsilon_{n} \begin{pmatrix}
        u_{n\uparrow}^i \\ u_{n\downarrow}^i \\ v_{n\uparrow}^i \\ v_{n\downarrow}^i
    \end{pmatrix} . \label{eq:BdG_eq}
\end{equation}
that are complimented by the self-consistency conditions
\begin{gather}
    \Delta_i = \frac{U}{2} \sum_{n}' (u_{n \uparrow}^i v_{n \downarrow}^{i*} - u_{n \downarrow}^i v_{n \uparrow}^{i*}), \quad \rho_{i\sigma\sigma'} = (2\delta_{\sigma\sigma'} - 1)\sum_{n}' v_{n \sigma}^i v_{n \sigma'}^{i*}.
\end{gather}

The solution to this problem will provide us with insights into the behavior of our system under the conditions described.

Throughout this work, we consider hexagonal samples with edge length $L = 30$ lattice constants. Periodic boundary conditions were applied. The stability of our solutions was verified against different random initial conditions, assuming convergence when the maximal difference of the chemical potential $\mu$ and the fields $\Delta_i$ and $\rho_{i\sigma\sigma'}$ at any lattice site between two consecutive iterations is less than a specified tolerance level $\alpha_{\text{tol}}$, set at $\alpha_{\text{tol}} = 10^{-5}$. The robustness of our solutions was further evaluated by varying $\alpha_{\text{tol}}$ and comparing the results.

\section{Analytical Solution Neglecting Self-Consistency\label{sec:NSC}}

In this section, we analytically derive the effect a weak magnetic field has on the energy of the Yu-Shiba-Rusinov state. For simplicity in our narrative, we will assume that $V_0$ (the non-magnetic part of the local potential) is zero, cf. Eq. (5) of the main text.

We begin by formulating the equation to determine the wave-functions $\bm{\psi}_n$ and energies $\varepsilon_n$
\begin{gather}
    \left(\xi_p [\sigma^0 \otimes \tau^z] + \Delta [\sigma^x \otimes \tau^x] -B [\sigma^z \otimes \tau^z] -J \hat{R}_{(\theta, \phi)}^{-1} [\sigma^z \otimes \tau^z] \hat{R}_{(\theta, \phi)} \delta(\bm{r}) \right) \bm{\psi}_n = 
    \varepsilon_n \bm{\psi}_n. \label{eq:EqBdG}
\end{gather}
Here, $\xi_p = \epsilon_p - \varepsilon_{\text{F}}$ represents the deviation of the energy spectrum $\epsilon_p$ of the electrons in the lattice from the Fermi level $\varepsilon_{\text{F}}$, and we assumed a homogeneous pairing potential $\Delta$.

For analytical convenience, we consider a rotated basis in which the magnetic moment of the impurity points in the $(0,0,1)$ direction. This is accomplished by introducing the rotation matrix $\hat{R}_{(\theta, \phi)}$, which, in the basis $(u_{\uparrow}, \; u_{\downarrow}, \; v_{\uparrow}, \; v_{\downarrow})^T$, reads as
\begin{align}
    \hat{R}_{(\theta, \phi)} = 
    \begin{pmatrix}
         \cos \frac{\theta}{2} & \sin \frac{\theta}{2} & 0 & 0 \\
        -\sin \frac{\theta}{2} & \cos \frac{\theta}{2} & 0 & 0 \\
        0 & 0 & \cos \frac{\theta}{2} &  \sin \frac{\theta}{2} \\
        0 & 0 & - \sin \frac{\theta}{2} &  \cos \frac{\theta}{2}
    \end{pmatrix} 
    \begin{pmatrix}
        e^{i\phi/2} & 0 & 0 & 0 \\
        0 & e^{-i\phi/2} & 0 & 0 \\
        0 & 0 & e^{-i\phi/2} & 0 \\
        0 & 0 & 0 & e^{i\phi/2}
    \end{pmatrix}.
    \label{eq:R_rot}
\end{align}
Upon applying this rotation to our formulation in Eq.~\eqref{eq:EqBdG}, we immediately note that only the component arising from the homogeneous magnetic field is altered. In the rotated basis, we find
\begin{equation}
    \hat{R}_{(\theta, \phi)} B [\sigma^z \otimes \tau^z] \hat{R}_{(\theta, \phi)}^{-1} = -B \begin{pmatrix}
        \cos \theta & -\sin \theta & 0 & 0 \\
        -\sin \theta & -\cos \theta & 0 & 0 \\
        0 & 0 & -\cos \theta & \sin \theta \\
        0 & 0 & \sin \theta & \cos \theta
    \end{pmatrix} \label{eq:H_M_rotated}.
\end{equation}

In the subsequent analysis, we will incorporate the magnetic field perturbatively. To that end, let us consider the unperturbed problem,
\begin{gather}
    \left(\xi_p [\sigma^0 \otimes \tau^z] + \Delta [\sigma^x \otimes \tau^x] -J [\sigma^z \otimes \tau^z] \delta(\bm{r}) \right) \bm{\psi}_n^{(0)} = 
    \varepsilon_n^{(0)} \bm{\psi}_n^{(0)} .\label{eq:EqBdG_noB}
\end{gather}
Switching to momentum space, we immediately obtain
\begin{gather}
    \bm{\psi}_{n}^{(0)}(\bm{p}) = J [\xi_p + \Delta - \varepsilon_n^{(0)}]^{-1} [\sigma^z \otimes \tau^z] \bm{\psi}_n^{(0)}(\bm{r} = 0).
\end{gather}
Upon reverting back to real space and seeking the solution at the impurity's site ($\bm{r} = 0$), we are presented with the following matrix equation:
\begin{equation}
    \bm{\psi}_n^{(0)}(\bm{r} = 0) = \frac{\alpha}{\sqrt{\Delta^2 - \varepsilon_n^{(0)2}}} \begin{pmatrix}
        \varepsilon_n^{(0)} & 0 & 0 & \Delta \\
        0 & -\varepsilon_n^{(0)} & \Delta & 0 \\
        0 & \Delta & -\varepsilon_n^{(0)} & 0 \\
        \Delta & 0 & 0 & \varepsilon_n^{(0)}
    \end{pmatrix} \bm{\psi}_n^{(0)}(\bm{r} = 0), \quad \alpha = \pi \nu_0 J .
\end{equation}
Here, $\nu_0$ denotes the density of states at the Fermi energy. Focusing on the in-gap state, where $\bm{\psi}_\text{YSR}^{(0)}(\bm{0}) \neq 0$ \footnote{Noteworthy, the magnetic impurity influences the above-the-gap continuum of states by altering the plane waves into their combinations, such that $\bm{\psi}_p^{(0), \pm}(\bm{0}) = 0$. Here, $\bm{\psi}_p^{(0), \pm}(\bm{r})$ is the wave-function corresponding to $\varepsilon_p^{(0),\pm} = \pm \sqrt{\xi_p^2 + \Delta^2}$.}, this yields (for $\alpha > 0$):
\begin{equation}
    \bm{\psi}_{\rm YSR}^{(0),+}(\bm{r} = 0) = \frac{1}{\sqrt{\mathcal{N}}}\begin{pmatrix}
        1 \\ 0 \\ 0 \\ 1
    \end{pmatrix}, 
    \quad \bm{\psi}_{\rm YSR}^{(0),-}(\bm{r} = 0) = \frac{1}{\sqrt{\mathcal{N}}}\begin{pmatrix}
        0 \\ 1 \\ 1 \\ 0
    \end{pmatrix}, \quad \varepsilon_\text{YSR}^{(0), \pm} = \pm \Delta \frac{1-\alpha^2}{1+\alpha^2} .
\end{equation}
and for $\alpha < 0$
\begin{equation}
    \bm{\psi}_{\rm YSR}^{(0), +}(\bm{r} = 0) = \frac{1}{\sqrt{\mathcal{N}}}\begin{pmatrix}
        0 \\ 1 \\ -1 \\ 0
    \end{pmatrix}, 
    \quad \bm{\psi}_{\rm YSR}^{(0), -}(\bm{r} = 0) = \frac{1}{\sqrt{\mathcal{N}}}\begin{pmatrix}
        1 \\ 0 \\ 0 \\ -1
    \end{pmatrix}, \quad \varepsilon_\text{YSR}^{(0), \pm} = \pm \Delta \frac{1-\alpha^2}{1+\alpha^2} .
\end{equation}
In the above expressions, $\mathcal{N}$ is the normalization constant that ensures $\int \frac{d^2\bm{p}}{(2\pi)^2} \psi_{\text{YSR}}^{(0), \pm} (\bm{p}) \psi_{\text{YSR}}^{(0), \pm} (-\bm{p})^* = 1$. It is given by
\begin{equation}
    \mathcal{N} = \frac{2 J \alpha \Delta}{\left(\Delta - |\varepsilon_{\rm YSR}^{(0), \pm}|\right) \sqrt{\Delta^2 - (\varepsilon_{\rm YSR}^{(0),\pm})^2}} = \frac{J (1+\alpha^2)^2}{2 \alpha^2 \Delta}. 
\end{equation}

If we now perturbatively incorporate the magnetic field as described by Eq.~\eqref{eq:H_M_rotated}, we readily find:
\begin{equation}
    \delta \varepsilon_\text{YSR}^{(1), \pm} = \mp B \cos \theta .
\end{equation}

In order to calculate the lowest order correction to the free energy $\delta E_g^{(1)}$, we recall that this correction can be conveniently formulated as
\begin{equation}
    \delta E_g^{(1)} = \langle \hat{H}_{\mathrm{M}} \rangle = - B \int d\bm{r} (\rho_{\uparrow \uparrow}^{(0)}(\bm{r}) - \rho_{\downarrow \downarrow}^{(0)}(\bm{r})) = - B m^{z, (0)},  \label{eq:H_M_1}  
\end{equation}
where $\rho_{\uparrow \uparrow}^{(0)}(\bm{r})$ and $\rho_{\downarrow \downarrow}^{(0)}(\bm{r})$ are the spin-resolved densities induced by the magnetic impurity in the absence of a magnetic field $B\bm{b}$. Eq.~\eqref{eq:H_M_1} suggests that the correction involves a Zeeman coupling of the magnetic field to the impurity-induced magnetization $m^{z, (0)} = \int d\bm{r} [\rho_{\uparrow \uparrow}^{(0)}(\bm{r}) - \rho_{\downarrow \downarrow}^{(0)}(\bm{r})]$, which assumes the value $0$ when $\alpha < \alpha_c^{(0)}(\beta)$ and $\cos \theta$ when $\alpha > \alpha_c^{(0)}(\beta)$.

\section{Ground State Energy in Weak Coupling Regime\label{sec:WeakCoupling}}

\begin{figure}[t!]
    \centering    \includegraphics[width=0.325\textwidth]{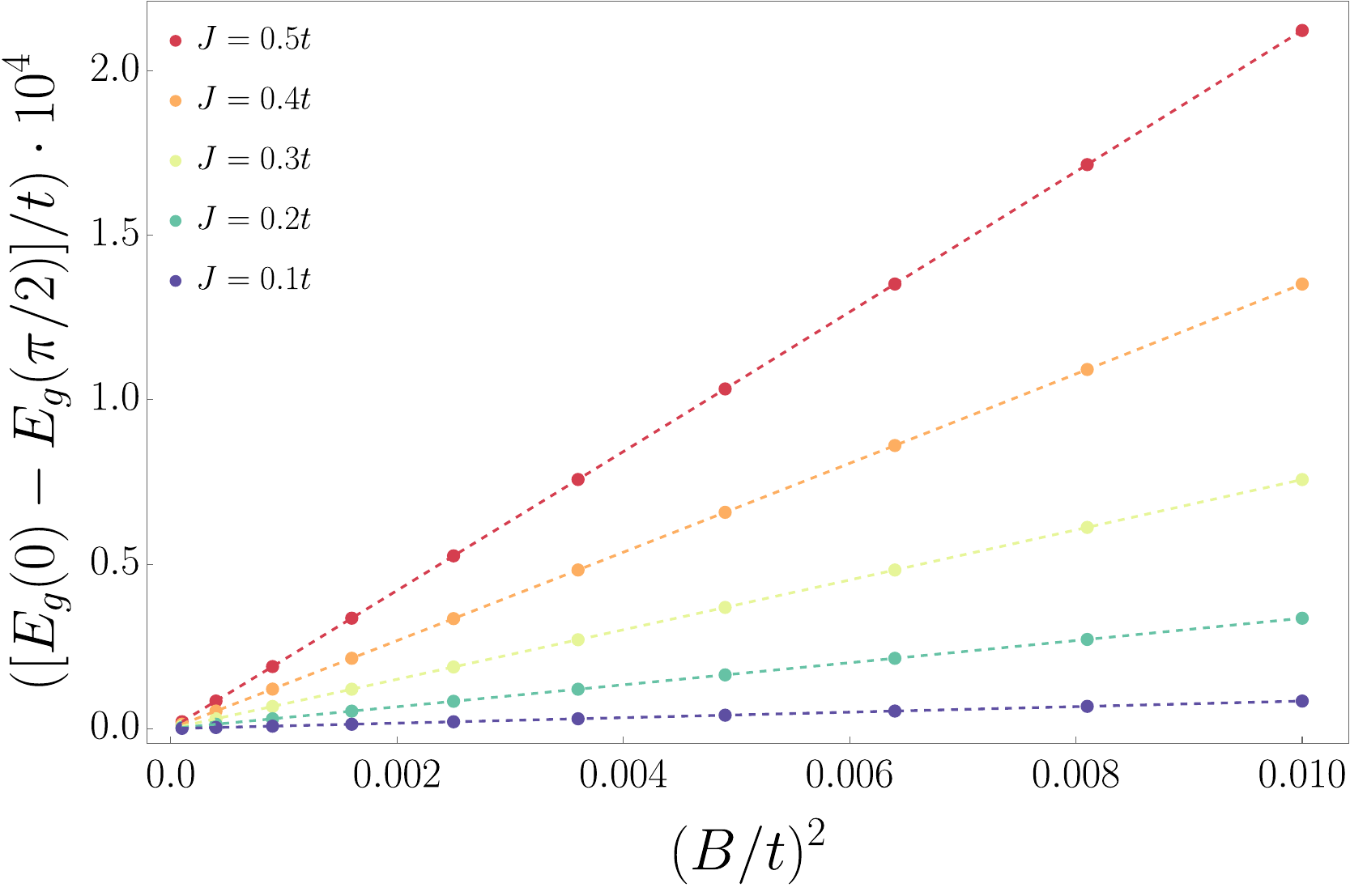}
    \includegraphics[width=0.325\textwidth]{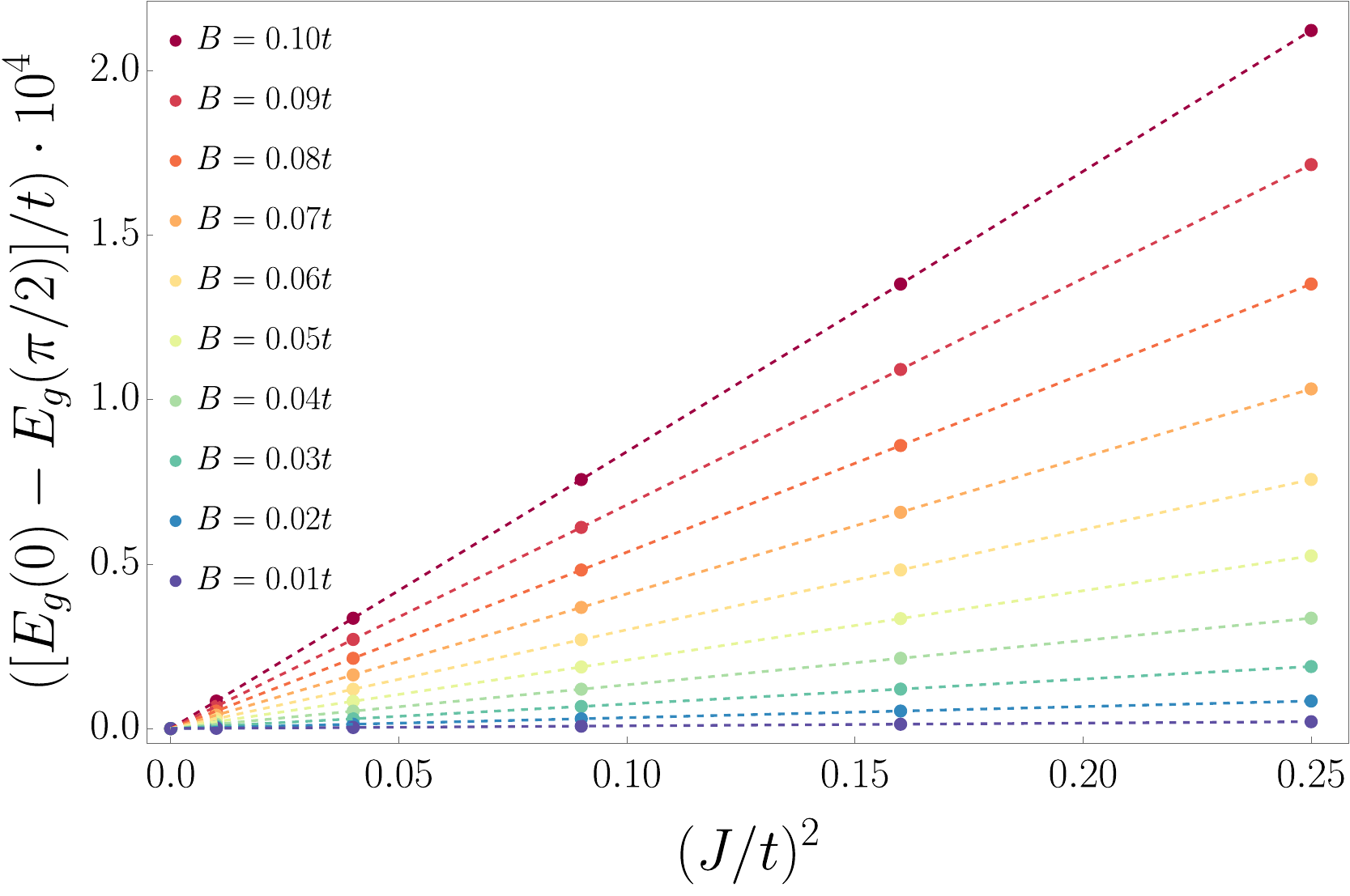}
    \includegraphics[width=0.334\textwidth]{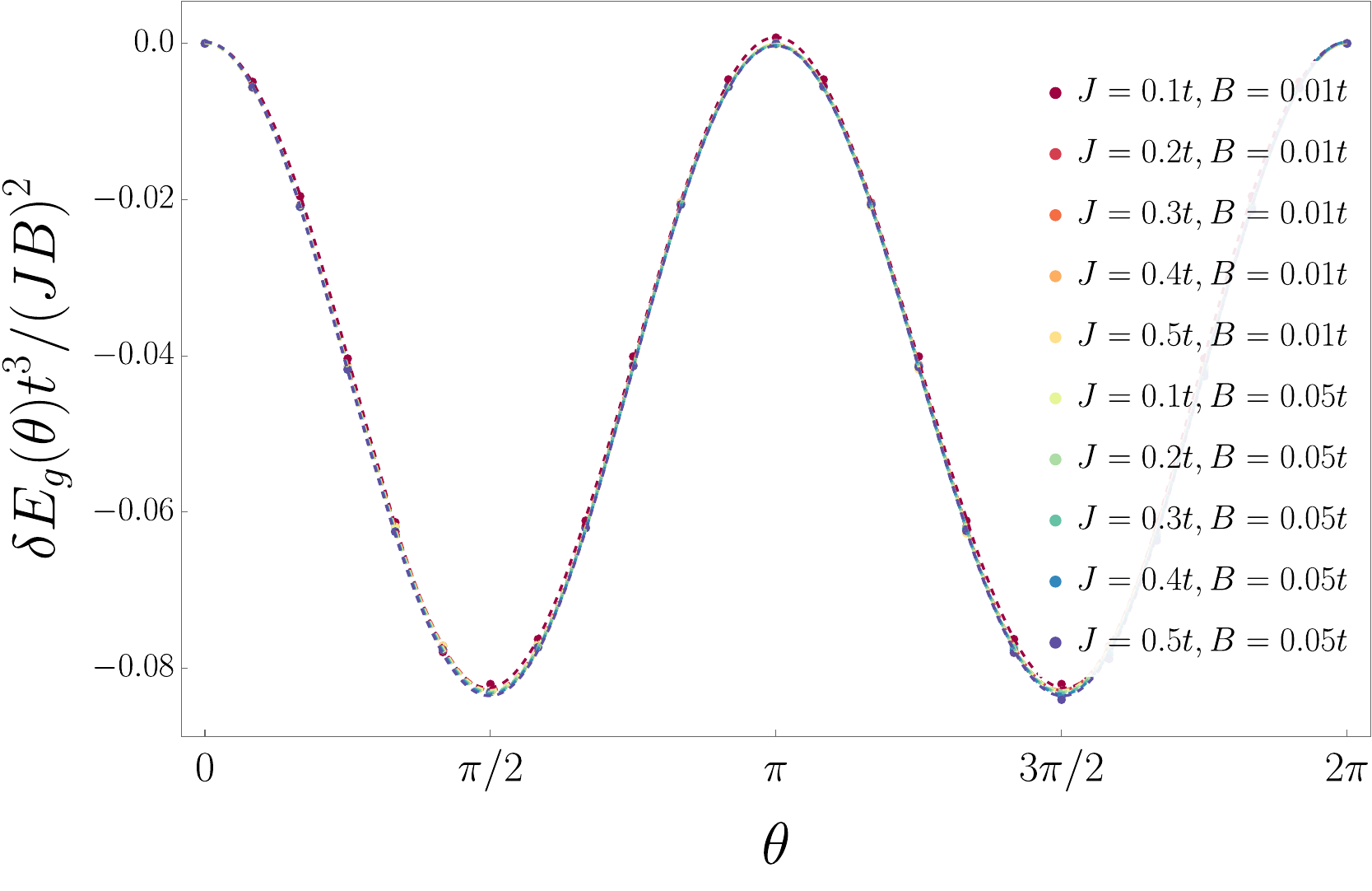}
    \caption{\textit{Left}: The dependence of the amplitude $[E_g(0) - E_g(\pi/2)]/t$ on $(B/t)^2$ for different values of the coupling strength $J$. The plot suggests linear behavior. \textit{Middle}: The behavior of $[E_g(0) - E_g(\pi/2)]/t$ as a function of $(J/t)^2$ for different values of the magnetic field $B$. Here again, linear plots are observed. \textit{Right}: Evolution of $\delta E_g(\theta) = E_g(\theta) - E_g(0)$ divided by $(JB)^2/t^3$ for various $J$ and $B$ values. The plots show remarkable agreement with one another, supporting the $(JB)^2$ behavior of the $\cos 2\theta$ harmonics. (Parameters: $n=1.0,  \Delta = 0.443t, V_0 = 0.0t, B=0.05t$)}
    \label{fig:E_g_weak}
\end{figure}

In this section, we present numerical results for the weak-coupling regime, where the net electronic spin is $0$ and, consequently, the ground state energy remains independent of $B$ at the first order, cf. Eq.~\eqref{eq:H_M_1}. Assuming a homogeneous gap $\Delta = 0.443t$, a filling factor of $n=1.0$, and $V_0 = 0.0t$, we explored the variation of the coupling strength $J$ and the magnitude of the external field $B$. This analysis revealed a $\cos 2\theta$ dependence of $E_g(\theta)$, as illustrated in the rightmost plot of Fig.~\ref{fig:E_g_weak}. To ascertain the parametric dependence of the amplitude of this harmonic, we examined $\ln[E_g(\theta = 0) - E_g(\theta = \pi/2)]$ with respect to both $\ln B$ and $\ln J$. Our analysis suggests a quadratic dependence on both $J$ and $B$, i.e., $A_2 (JB)^2 \cos 2\theta$ with some constant $A_2$. For clarity of presentation, the behavior of $[E_g(0) - E_g(\pi/2)]/t$ as a function of $(B/t)^2$ and $(J/t)^2$ is depicted in the left and middle figures of Fig.~\ref{fig:E_g_weak}. To further substantiate our claim, we present the rightmost plot in Fig.~\ref{fig:E_g_weak}, that illustrates the evolution of $\delta E_g(\theta) = E_g(\theta) - E_g(0)$ normalized by $(JB)^2/t^3$, which also suggests that the amplitude of $E_g(\theta)$ is directly proportional to $(J B)^2$. We would like to emphasize that to render the scales dimensionless, in Fig.~\ref{fig:E_g_weak}, we normalized all quantities with respect to the hopping amplitude $t$. Nonetheless, the appropriate dimensionless parameters should be $\alpha = \pi \nu_0 J$ and $\gamma = \pi \nu_0 B$, which depend not only on $t$, but on the filling factor $n$ as well.

\section{Order Parameter and Magnetic Moment at the Impurity Site\label{sec:imp_site}}

This section presents typical plots, illustrated in Fig.~\ref{fig:impurity}, of the behavior of the local pairing potential $\Delta(r = 0)$ (normalized to the homogeneous value $\Delta_0$ far from the impurity) and the magnetic polarization $\bm{m}(r = 0)$ at the impurity's site ($r = 0$) as functions of the impurity-to-substrate coupling constant $J$. 

\begin{figure}[h!]
    \centering    \includegraphics[width=0.45\textwidth]{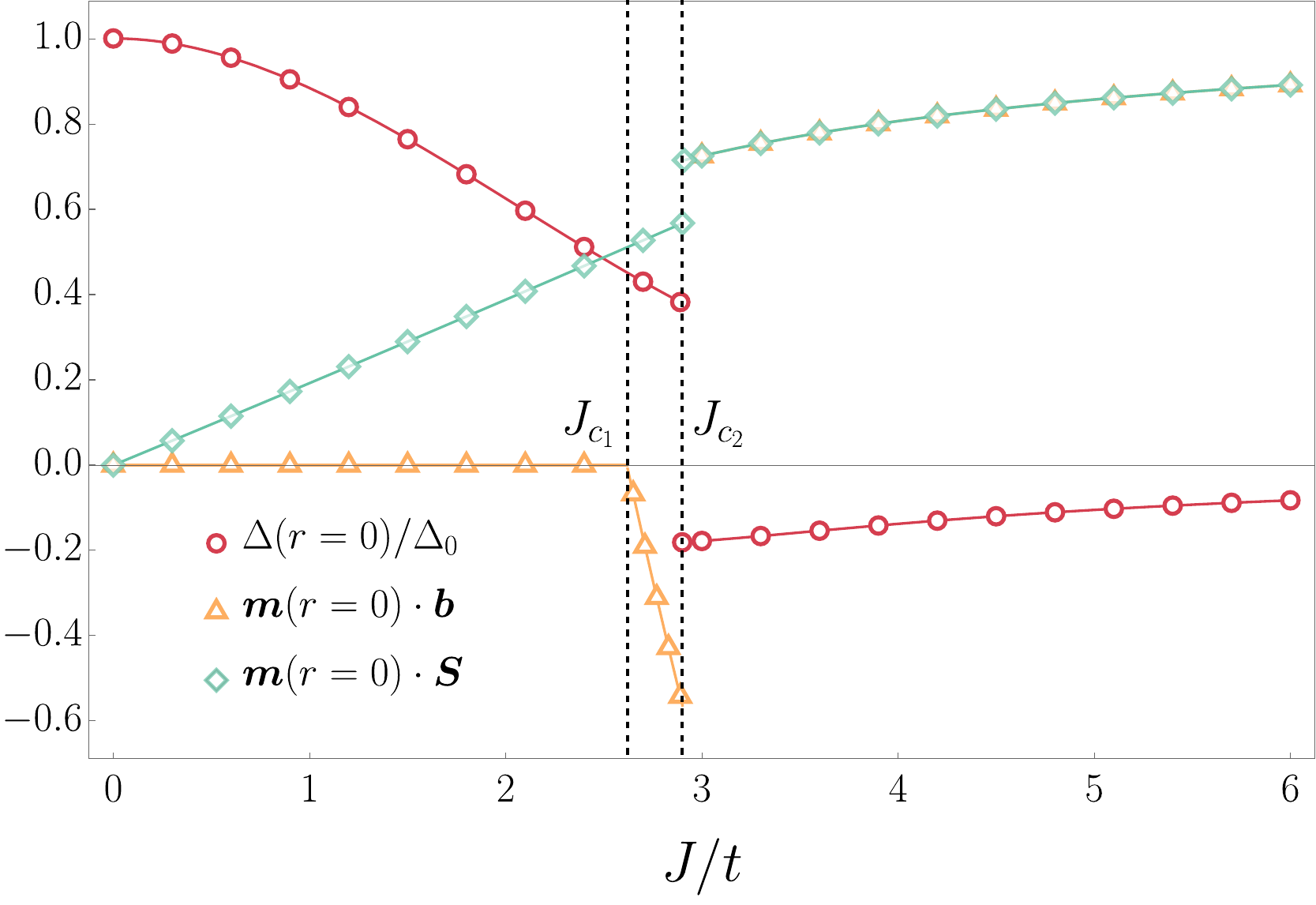}
    \caption{Evolution of $\Delta(r=0)/\Delta_0$ and $\bm{m}(r=0)$ at the impurity's site $r=0$ with the coupling strength $J$ measured in units of the hopping amplitude $t$. 
    Critical values $J_{c_1} \simeq 2.6t$ and $J_{c_2} \simeq 2.9t$ are marked with dashed lines. In the region $J_{c_1} < J < J_{c_2}$, the sign of $\bm{m}(r=0) \cdot \bm{b}$ is negative and $\varepsilon_{\rm YSR} = 0$, i.e. it adheres the Fermi level. (Parameters: $V_0 = 0.0t$, $B = 0.05t$)}
    \label{fig:impurity}
\end{figure}

Two critical values, $J_{c_1} \simeq 2.6t$ and $J_{c_2} \simeq 2.9t$, delineate the region where the YSR state's energy matches the Fermi level and the impurity's magnetic moment points in the opposite direction to the external magnetic field, forming an angle $\pi/2 < \theta_{\text{opt}}(J) \leqslant \pi$, cf. Fig. 1 of the main text. These thresholds are indicated by vertical dashed lines. It becomes further evident from this plot that within the region $J_{c_1} \leqslant J \leqslant J_{c_2}$, the magnetic moment of the impurity on the $\bm{b}$-axis (axis of the external magnetic field) assumes negative values. After crossing $J_{c_2}$, both $\Delta(r = 0)$, $\bm{m}(r = 0) \cdot \bm{S}$, and $\bm{m}(r = 0) \cdot \bm{b}$ exhibit abrupt jumps corresponding to the local destruction of Cooper pairs and alteration of the Fermi sea structure.

We note that $\delta \alpha = \pi\nu_0 (J_{c_2} - J_{c_1}) \simeq 0.1t \simeq B/\Delta$, which aligns with our analytical predictions, see Eq.~(18) of the main text.

\end{widetext}

\end{document}